\documentclass[%
 reprint,twocolumn,
 amsmath,amssymb,
 aps,pre
]{revtex4}
\usepackage{xcolor}
\usepackage{bbold}
\usepackage{graphicx}
\usepackage{dcolumn}
\newcommand{\compconj}[1]{%
  \overline{#1}%
}
\usepackage{bm}

\begin{document}

\preprint{APS/123-QED}

\title{Heat rectification, heat fluxes, and spectral matching }

\author{Javier Navarro}
\affiliation{%
Department of Physical Chemistry, University of the Basque Country UPV/EHU, Apdo 644, Bilbao, Spain
}%


\author{Juan Gonzalo Muga}
\affiliation{
Department of Physical Chemistry, University of the Basque Country UPV/EHU, Apdo 644, Bilbao, Spain
}%
\affiliation{
EHU Quantum Center, University of the Basque Country UPV/EHU, 48940 Leioa, Spain
}%
\author{Marisa Pons}
\affiliation{%
 Dapartament of Applied Physics, University of the Basque Country UPV/EHU, Bilbao, Spain}
\affiliation{ EHU Quantum Center, University of the Basque Country UPV/EHU, 48940 Leioa, Spain
}%


\date{\today}

\begin{abstract}
Heat rectifiers would facilitate energy management operations such as cooling, or energy harvesting, but
devices of practical interest are still missing.
Understanding heat rectification at a fundamental level is
key to help us find or design such devices.
The match or mismatch of the phonon band spectrum of device segments
for forward or reverse temperature bias of the thermal baths at device boundaries, was
proposed as the mechanism behind
rectification. However no explicit, theoretical  relation derived from first principles
had been found so far between  heat fluxes and spectral matching.
We study heat rectification in a minimalistic chain of two coupled ions.
The fluxes and rectification can be calculated analytically.
We propose a definition of the matching that sets an upper bound for the
heat flux. In a regime where the device rectifies optimally,  matching and flux ratios for forward and reverse
configurations are
found to be proportional.
The results can be extended to a system of $N$ particles in arbitrary traps with nearest-neighbor linear interactions.
\end{abstract}
%
\maketitle
%
%
%
%
\section{Introduction \label{sec:Introduction}}
Heat rectification is a phenomenon in which the thermal energy that flows through a device between two reservoirs
depends on the sign of their temperature bias [\onlinecite{Roberts2011,Li,Pereira2019}].
Thus, an ideal heat rectifier or thermal diode would let heat flow only in one direction, for the ``forward bias'',
and act as an insulator for the ``reverse bias'' configuration with the bath temperatures exchanged.
Such devices would serve for different energy management and thermal control operations, such as energy harvesting, refrigeration,
or to implement thermal-based transistors, logic gates and logic circuits \cite{Li, Wang, transist}. Proposed physical platforms for their applications go from the macro \cite{Roberts2011} to the microscale, for example in nanostructures \cite{Ma2019}, or trapped ions \cite{S2019,S2021}.
The first experimental observations of this interesting phenomenon were due to Starr in 1936 [\onlinecite{Starr}].
Since then, much work has been done,
but we are far from achieving useful devices \cite{Chen2015,Pereira2019} in spite of the exploration of many different factors such as
surface roughness/flatness at material contacts \cite{Roberts2011}, thermal potential barriers [\onlinecite{JMoon}], temperature dependence of thermal conductivity between different materials [\onlinecite{Marucha}], nanostructured asymmetry (i.e. mass-loaded nanotubes, asymmetric geometries in nanostructures, nanostructured interfaces) [\onlinecite{Alaghemandi_2009}], anharmonic lattices \cite{Terra,Lucianno2021}, graded materials \cite{Wang2012}, long range interactions \cite{Pereira2013},
localized impurities \cite{Marisa,Alexander2020},
or quantum effects
[\onlinecite{PhysRevLett.97.094301,Pereira2019}].
For a more extensive list of references see the reviews \cite{Roberts2011,Li,Pereira2019,Ma2019}.

Theoretical work started with Terraneo {\it{et al.}} [\onlinecite{Terra}].
They showed thermal rectification in  a segmented chain of coupled nonlinear oscillators in contact
with two thermal baths at different temperatures.
The heat rectification was understood as a consequence of
the match or mismatch of the phonon spectra of the different segments of the 1D chain when changing the temperature bias  \cite{Terra,ITRLi,Li1, Li}.
The different dependences of the segments spectra with respect to temperature,
implied conduction or isolation for the forward or the reverse bias.  Li {\it{et al.}} \cite{ITRLi},
to describe the efficiency of the rectifier,
analyzed the ratio of heat fluxes $|J/\tilde{J}|$ between the forward, $J$, and reverse, $\tilde{J}$, configurations.
They found numerically, for their coupled nonlinear lattices model, a logarithmic relation between this ratio
and the ratio of the degrees of overlap $|J/\tilde{J}|^\delta \sim {\cal{S}}/\tilde{\cal S}$ with ${\cal S}$
and $\tilde{\cal S}$ being measures of the phonon-band overlap in the forward and reverse configurations. Yet
this relation was not inferred from first principles. A theoretical connection between flux and matching, beyond the numerical findings,
has been missing so far.


Nonlinear forces in the chain result in a temperature dependence of the phonon bands or power spectrum densities,
possibly leading to rectification.
However Pereira \cite{Pereira} pointed out that nonlinear forces are not a necessary condition for rectification,
which only needs some structural asymmetry and a temperature dependence of some system parameters to occur.
Indeed, the linear regime (i.e. harmonic interactions) is quite natural and realistic in some systems, such as trapped ions. Heat transport
in trapped ion chains has been studied in several works
\cite{PhysRevE.99.062105, PhysRevB.89.214305, Pruttivarasin_2011,Freitas_2015}.
Sim\'on {\it{et al.}} \cite{S2019,S2021}  proposed trapped ions as an experimentally feasible setting for heat rectification.
They numerically demonstrated first heat rectification for linear chains of ions
with graded trapping frequencies \cite{S2019}, and later in a minimalistic two-ion model \cite{S2021}.
For two trapped ions the asymmetry may be provided by different species
and the effective baths are implemented by Doppler cooling lasers that imply a temperature dependence of the couplings.
The model is also quite interesting because the analytical treatment of several quantities,
such as the flux, allows us to find optimal rectification conditions \cite{S2021}.
Moreover
trapped ions constitute a well-developed and tested architecture for
fundamental research, quantum information processing, and
quantum technologies such as detectors or metrology.  This
architecture is in principle scalable in driven ion circuits (see, e.g., \cite{S21}). Controllable heat rectification in this context
would be a useful asset for energy management in trapped-ion based technologies.

In this paper we find, for the two-ion linear ion chain, that a properly defined matching of the
phononic spectra is an upper bound for the thermal flux.
In Sect. \ref{sec:Physical_Model} we provide an overview of the model. In Sect. \ref{sec:FluxMatching}, we find a
general relation between the thermal flux and the matching of the spectral densities. In Sect. \ref{sec:NumRes}
matching and the flux are compared numerically. Finally, in Sect. \ref{sec:conclusions} we present the conclusions and a generalization.
\section{PHYSICAL MODEL \label{sec:Physical_Model}}
\begin{figure}
	\center
   \includegraphics[scale=0.57]{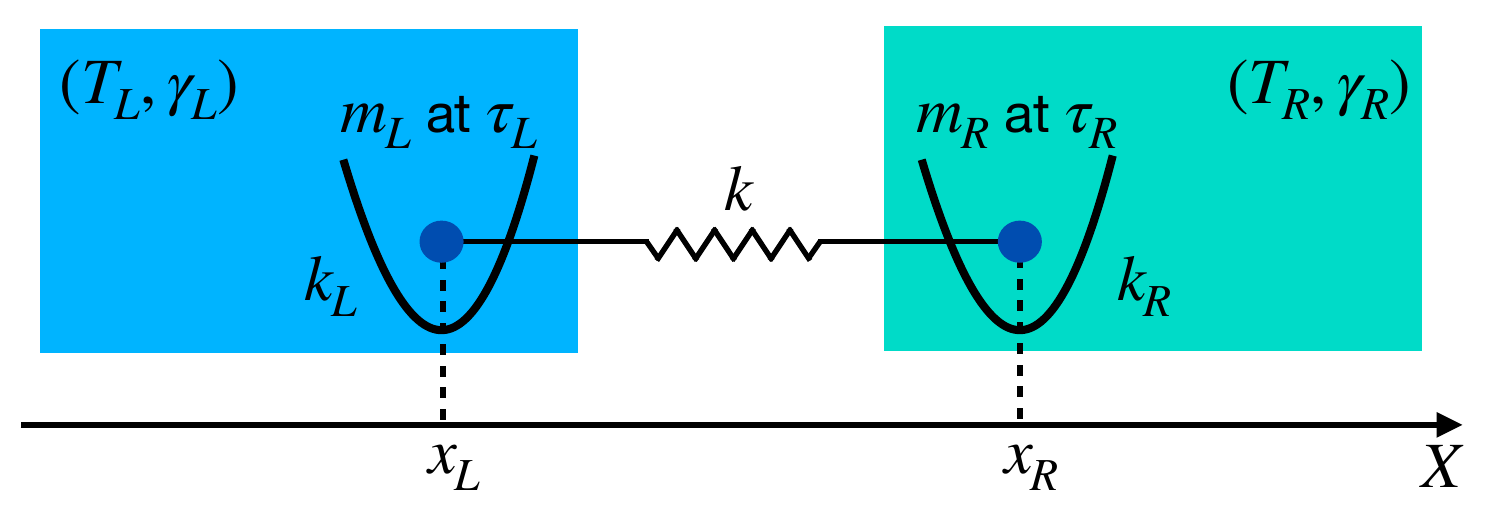}
	\caption {Scheme of the model described in Sec. \ref{sec:Physical_Model}. Each mass $m_i$ is at temperature $\tau_i$, ($i=L,R$ is the generic index for ``left'' or ``right''),   trapped by a harmonic potential, and connected to a thermal bath at a temperature $T_i$  and friction coefficient $\gamma_i$.
The masses interact by a harmonic potential with each other.}
	\label{FIGmodel}
\end{figure}

The minimalistic two-ion model describes two ions in individual traps subjected to Doppler cooling lasers \cite{S2021} and a mutual Coulomb interaction, see Fig. \ref{FIGmodel}.   In the small oscillations regime, which is realistic for ions in multisegmented Paul traps,
the model boils down mathematically to two harmonically coupled
masses $m_L$ and $m_R$ (the subscripts refer to left and right and, when needed, will be described generically by the index $i=R,L$). Each mass is confined into a harmonic potential with spring constants $k_L$ and $k_R$
respectively, and in contact with thermal baths at different temperatures, $T_L$ and $T_R$. The two masses are
coupled through a spring with constant $k$ \cite{S2021}. $x_L$ is the position of mass $m_L$ and
$x_R$ is the position of mass $m_R$.

Without any coupling to the baths
the system Hamiltonian is
\begin{equation}
H=\frac{p_{L}^{2}}{2 m_L}+\frac{p_{R}^{2}}{2 m_R}+V\left(x_L, x_R\right),
\label{b2}
\end{equation}
with $\quad V\left(x_L, x_R\right)=$  
 $[{k_L}\left(x_L-x_{eL}\right)^2+$  ${k_R}\left(x_R-x_{eR}\right)^2+
{k}\left(x_L-x_R-x_e\right)^2]/2$, where $\left\{x_{i}, p_{i}\right\}_{i=L,R}$ are the position and momentum of each mass, $x_{eL}$ is the center of the left ion trap, $x_{eR}$ is the center of the right ion trap, and $x_e$ is the natural length of the linear coupling. Changing  coordinates to the displacements from equilibrium positions of the system, $q_{i}=x_{i}-x_{i}^{e q}$, where $x_{i}^{e q}$ are the solutions to $\partial_{x_{i}} V\left(x_{L}, x_{R}\right)=0$, the Hamiltonian can be written as
\begin{align}
\begin{split}
\label{h1}
    H&=\frac{p_{L}^{2}}{2 m_L}+\frac{p_{R}^{2}}{2 m_R}+\frac{k+k_{L}}{2} q_{L}^{2} \\
    &+\frac{k+k_{R}}{2} q_{R}^{2}-k q_{L} q_{R}+V\left(x_{L}^{e q}, x_{R}^{e q}\right) .
\end{split}
\end{align}
For later use let us define $V_i=({k+k_{i}})q_{i}^{2}/2$ and $V_{LR}=-k q_{L} q_{R}$.
The constant term $V\left(x_{L}^{e q}, x_{R}^{e q}\right)$ does not affect the evolution of the system so it can be ignored.
The baths are modeled as Langevin baths so the friction coefficients $\gamma_L$, $\gamma_R$, and the Gaussian white-noise-like forces $\xi_L$ and $\xi_R$ are introduced into the equations of motion,
\begin{equation}
\begin{split}
\dot{q}_{L}&=\frac{p_{L}}{m_L}, \quad \dot{q}_{R}=\frac{p_{R}}{m_R} ,\\
\dot{p}_{L}&=-\left(k+k_{L}\right) q_{L}+k q_{R}-\frac{\gamma_{L}}{m_L} p_{L}+\xi_{L}(t), \\
\dot{p}_{R}&=-\left(k+k_{R}\right) q_{R}+k q_{L}-\frac{\gamma_{R}}{m_R} p_{R}+\xi_{R}(t),
\label{motion}
\end{split}
\end{equation}
where the following averages over noise realizations are assumed: $\langle \xi_i (t)\rangle=0$, $\left\langle\xi_{L}(t) \xi_{R}\left(t^{\prime}\right)\right\rangle=0$, $\left\langle\xi_{L}(t) \xi_{L}\left(t^{\prime}\right)\right\rangle=2 D_{L} \delta\left(t-t^{\prime}\right)$, and $\left\langle\xi_{R}(t) \xi_{R}\left(t^{\prime}\right)\right\rangle=2 D_{R} \delta\left(t-t^{\prime}\right)$. The diffusion coefficients $D_{L}$ and $D_{R}$ obey $D_{L}=\gamma_{L} k_{B} T_{L} $ and $D_{R}=\gamma_{R} k_{B} T_{R}$, where $k_B$ is Boltman's constant.

A compact notation for the equations of motion is
\begin{equation}
\dot{\vec{r}}(t)=\mathbb{A} \vec{r}(t)+\mathbb{L} \vec{\xi}(t),
\label{eqm}
\end{equation}
where
$\vec{r}(t)\equiv\left(\vec{q}, \mathbb{M}^{-1} \vec{p}\right)^{\top}=(q_L,q_R,\dot{q_L},\dot{q_R})^\top$ (the superscript $\top$ means ``transpose''), $\mathbb{M}=\operatorname{diag}\left(m_L, m_R\right)$, and
%
\begin{align}
\begin{split}
\mathbb{A}&=\left(\begin{array}{cc}
\mathbb{0}_{2 \times 2} & \mathbb{1}_{2 \times 2} \\
-\mathbb{M}^{-1} \mathbb{K} & -\mathbb{M}^{-1} \mathbb{\Gamma}
\end{array}\right), \\
\mathbb{L}&=\left(\begin{array}{l}
\mathbb{0}_{2 \times 2}, \\
\mathbb{M}^{-1}
\end{array}\right),\\
\mathbb{K}&=\left(\begin{array}{cc}
\frac{k+k_L}{2} & -k \\
-k& \frac{k+k_R}{2}
\end{array}\right).
\end{split}
\end{align}
Also $\vec{\xi}(t)=\left(\xi_{L}(t), \xi_{R}(t)\right)^{\top}$ (note that $\mathbb{L}$ is a $4\times 2$ matrix), $\mathbb{\Gamma}=\operatorname{diag}\left(\gamma_{L}, \gamma_{R}\right)$, $ \mathbb{0}_{2 \times 2}$ is the $2 \times 2$ matrix with all the components 0, and $\mathbb{1}_{2 \times 2}$ is the $2 \times 2$ identity matrix.


The baths are implemented by optical molasses (Doppler cooling lasers)
which set an effective temperature for each bath $T (=T_L,T_R)$, and an effective friction coefficient $\gamma(=\gamma_L,\gamma_R)$ which are controlled with the laser intensity $I$ and frequency detuning $\delta$ with respect to the selected internal atomic transition,
\begin{align}
  \begin{split}
\gamma(I, \delta) & =-4 \hbar\left(\frac{\delta+\omega_0}{c}\right)^2\left(\frac{I}{I_0}\right) \frac{2 \delta / \Gamma}{\left[1+(2 \delta / \Gamma)^2\right]^2},
 \\
T(\delta) & =-\frac{\hbar \Gamma}{4 k_B} \frac{1+(2 \delta / \Gamma)^2}{(2 \delta / \Gamma)},
\label{tgamma}
\end{split}
\end{align}
where $\omega_0$ is the (angular) frequency of the transition, $c$ is the speed of light, $I_0$ is the saturation intensity, and $\Gamma$ is the decay rate of the excited state. If $\Gamma$ and $I$ are fixed, $\gamma$ depends on $\delta$, and thus, indirectly, on the temperature $T$.
In the two-ion model we deal in general with two different species which involve two different atomic transitions, so the laser wavelengths and the decay rates  $\Gamma$ depend on the species. Then, exchanging the temperatures by modifying the detunings, keeping the laser intensities constant, does not necessarily imply an exchange of the friction coefficients. Nevertheless, it is possible to adjust the laser intensities so that the friction coefficients get exchanged and this is the assumption in \cite{S2021} and hereafter.
\subsection{Covariance and spectral density}
We are mostly interested in quantities such as the fluxes or particle temperatures in the steady state (s.s.) regime that is achieved after
sufficiently long time. These quantities can be computed from the ``marginal''
correlation matrix ${\mathbb{P}}^{s.s}=\langle \vec{r}(t) \vec{r}^\top(t)\rangle_{s.s}$, which in the stationary regime does not depend on $t$.

Using the steady state condition,
and Novikov's theorem,  the marginal covariance matrix in the steady state obeys \cite{S2019,Wiener}
\begin{equation}
\mathbb{A} \mathbb{P}^{\text {s.s. }}+\mathbb{P}^{\text {s.s. }} \mathbb{A}^{\top}=-2 \mathbb{L} \mathbb{D} \mathbb{L}^{\top},
\label{a7}
\end{equation}
where $\mathbb{D}=\operatorname{diag}(D_L, D_R )$.
This equation may be used to solve for $\mathbb{P}^{s.s}$.
Alternatively the Fourier space may also be used.
$\mathbb{P}^{s.s}=\mathbb{C}^{s . s .}(0)$ is a particular case ($\tau=0$) of the
steady-state covariance matrix $\mathbb{C}^{s . s .}(\tau)=\langle \vec{r}(t) \vec{r}^\top(t+\tau)\rangle_{s.s}$,
which, according to the Wiener-Khinchin theorem \cite{Wiener}
\begin{equation}
\mathbb{C}^{s.s}(\tau)=\left\langle\vec{r}(t) \vec{r}^{\top}(t+\tau)\right\rangle_{s.s}=\mathcal{F}^{-1}\left[\mathbb{S}_{\vec{r}}(\omega)\right](\tau),
\end{equation}
is the inverse Fourier transform of the
the spectral density matrix
\begin{equation}
\mathbb{S}_{\vec{r}}(\omega) \equiv\left\langle\vec{R}(\omega) \vec{R}^{\top}(-\omega)\right\rangle,
\end{equation}
where $\vec{R}(\omega)=(X_L,X_R,W_L,W_R)^\top$ is the Fourier transform (vector) of $\vec{r}$, namely
\begin{equation}
\mathbb{P}^{s.s}=\mathbb{C}^{s . s .}(0)=\frac{1}{2 \pi} \int_{-\infty}^{\infty}  \mathbb{S}_{\vec{r}}(\omega)\:d \omega.
\label{pcs}
\end{equation}
$\mathbb{S}_{\vec{r}}(\omega)$ may be computed as
\begin{equation}
\mathbb{S}_{\vec{r}}(\omega)=2(\mathbb{A}-i \omega)^{-1} \mathbb{L} \mathbb{D} \mathbb{L}^{\top}(\mathbb{A}+i \omega)^{-\top},
\label{sgen}
\end{equation}
see \cite{S2021,Wiener} for further details.

The diagonal matrix elements will be quite relevant for the analysis of the flux and matching.
In particular the  spectral densities ${S}_L \equiv \mathbb{S}_{3,3}(\omega)=\left\langle W_L(\omega) W_L(-\omega)\right\rangle$ for the left ion and ${S}_R \equiv \mathbb{S}_{4,4}(\omega)=\left\langle W_R(\omega) W_R(-\omega)\right\rangle$ for the right ion, where $W_i(\omega)$ is the Fourier transform of $\dot{q}_i(t)$, $i=L,R$, are (proportional to) power spectral densities of the kinetic energies since
$m_i\langle \dot{q}_i\dot{q}_i\rangle/2=m_i\int S_i d\omega/(4\pi)$, see Eq. (\ref{pcs}).
The spectral densities $S_L$ and $S_R$ in terms of velocity transforms are related to the
other diagonal elements $\mathbb{S}_{1,1}(\omega)$ and $\mathbb{S}_{2,2}(\omega)$, given in terms of displacement transforms,
using the Fourier transform of the derivative,
%
\begin{eqnarray}
    {S}_L&=&\omega^2\langle X_L(\omega)X_L(-\omega)\rangle,
    \nonumber\\
    {S}_R&=&\omega^2\langle X_R(\omega)X_R(-\omega)\rangle,
  \label{Ss}
\end{eqnarray}
a property that we shall use later on to relate spectral overlap and flux.
\subsection{Expressions for the flux}
We will find now expressions for the flux, starting with the
local
energy for the left particle, defined as
\begin{equation}
H_{L}=\frac{1}{2 m_L} p_{L}^{2}+V_L(q_L)+\frac{1}{2}  V_{LR}\left(q_{L},q_{R}\right).
\label{eq13}
\end{equation}
%
%
Differentiating with respect to time, we find the continuity equation
%
\begin{equation}
  \begin{split}
    \dot{H}_{L}=\frac{ p_{L}\dot{p_{L}}}{m_L}&+\frac{dV_L(q_L)}{dq_L}\dot{q_L}+\frac{1}{2}
    \frac{\partial V_{LR}\left(q_{L},q_{R}\right)}{\partial q_L}\dot{q}_L
    \\
    &+\frac{1}{2} \frac{\partial V_{LR}\left(q_{L},q_{R}\right)}{\partial q_R}\dot{q}_R.
    \label{a1}
\end{split}
\end{equation}
Using the equations of motion (\ref{eqm})
%
%
into Eq. (\ref{a1}), and simplifying, we get
\begin{equation}
  \begin{split}
\dot{H}_{L}=&\frac{ p_{L}}{m_L}F_{ext}-\frac{1}{2} \frac{\partial V_{LR}\left(q_{L},q_{R}\right)}{\partial q_L}\dot{q}_L\\+&\frac{1}{2} \frac{\partial V_{LR}\left(q_{L},q_{R}\right)}{\partial q_R}\dot{q}_R.
  \end{split}
  \label{hdot}
\end{equation}
where $F_{ext}=-\frac{\gamma_{L}}{m_L} p_{L}+\xi_{L}(t)$ includes the dissipative and the stochastic contributions.
The first term in Eq. (\ref{hdot}) due to the external force is the incoming flux of energy from the bath $J_{in}={ p_{L}}F_{ext}/m_L$. The second and third terms are the energy flux from particle $L$ to particle $R$,
\begin{equation}
\label{h2}
J_{LR}=-\frac{1}{2} \frac{\partial V_{LR}\left(q_{L},q_{R}\right)}{\partial q_L}\dot{q}_L+\frac{1}{2} \frac{\partial V_{LR}\left(q_{L},q_{R}\right)}{\partial q_R}\dot{q}_R.
\end{equation}
In the steady state $\langle \dot{H}_{L} \rangle=0$ so the incoming flux and the flux of energy from the left particle to the right particle obey $-\langle J_{in} \rangle= \langle J_{LR} \rangle$.
Then, the steady-state flux can be computed in two different ways.
We will calculate $\langle J_{LR} \rangle$ first.
Substituting $V_{LR}=-kq_Lq_R$ in Eq. (\ref{h2}),
\begin{equation}
J_{LR}=\frac{1}{2}k(q_R\dot{q}_L-q_L\dot{q}_R).
\label{a2}
\end{equation}
Since we are interested in average values we define
\begin{equation}
J=-\langle J_{LR}\rangle=\frac{k}{2}\left[\langle q_L\dot{q}_R\rangle  - \langle (q_R\dot{q}_L)\rangle\right].
\label{15}
\end{equation}
%

%

%
%
We apply the Wiener-Khinchin theorem to (\ref{15}), to find the heat flux in the steady state,
\begin{align*}
    J &=\frac{k}{4\pi} \int \left[\langle X_L(\omega)W_R(-\omega)\rangle- \langle X_R(\omega)W_L(-\omega)\rangle \right]d\omega\\
    &=\frac{ki}{4\pi}\!\!\int\!\! \omega \left[ \langle (X_R(\omega)X_L(-\omega))\rangle-\langle X_L(\omega)X_R(-\omega)\rangle\right]\!d\omega\\
    &=\frac{-k}{2\pi}\int \omega \:\operatorname{Im}\left[ \langle (X_R(\omega)X_L(-\omega))\rangle\right]d\omega,
    \stepcounter{equation}\tag{\theequation}\label{elflujo1}
\end{align*}
where, in the second line we have used the Fourier transform property ${W}_i(\omega)=i\omega X_i(\omega)$. Since the positions are real,
$X_i(\omega)=\compconj{X_i(-\omega)}$.

An alternative expression for the flux may be computed from the incoming flux,
\begin{equation}
  J_{in}=\frac{ p_{L}}{m_L}\left[-\frac{\gamma_{L}}{m_L} p_{L}+\xi_{L}(t) \right].
\end{equation}
Averaging,
\begin{equation}
  \langle J_{in}\rangle=-\frac{ \langle p_{L}^2\rangle}{m_L^2}\gamma_{L}+\frac{\langle p_{L}\xi_{L}(t)\rangle}{m_L}.
  \label{jk}
\end{equation}
Since the left particle temperature is
\begin{equation}
  \begin{split}
  &\tau_L(t)=\frac{\left\langle p_L^2(t)\right\rangle}{m_L k_B},
  \label{v1}
\end{split}
\end{equation}
Eq. (\ref{jk}) and Novikov's theorem, see  \cite{S2019,S2021} for a full calculation, give
\begin{equation}
    \langle J_{in}\rangle=k_B \frac{\gamma_L}{m_L}\left(T_L-\tau_L\right).
    \label{Js}
\end{equation}
For the steady state, $J=\langle J_{in}\rangle$ is equal to the alternative expression (\ref{elflujo1}).
%
%
%
%
%
\subsection{Rectification\label{recti}}
We use as a measure of rectification
the coefficient
%
\begin{equation}
R=\frac{|| J|-| \tilde{J}||}{\max (|J|,|\tilde{J}|)},
\label{rdefi}
\end{equation}
which is bounded between 0 and 1, $0\leq R\leq1$ .
%
Keep in mind that
to exchange the baths from forward to reverse bias implies here to exchange the temperatures and the friction coefficients.
A parametric exploration was done over the space formed by the parameters of the model $m_L$, $m_R$, $k$, $k_L$, $k_R$, $\gamma_L$ and $\gamma_R$ to maximize $R$ \cite{S2021}.
%
%
\begin{figure}
	\center
   \includegraphics[scale=0.45]{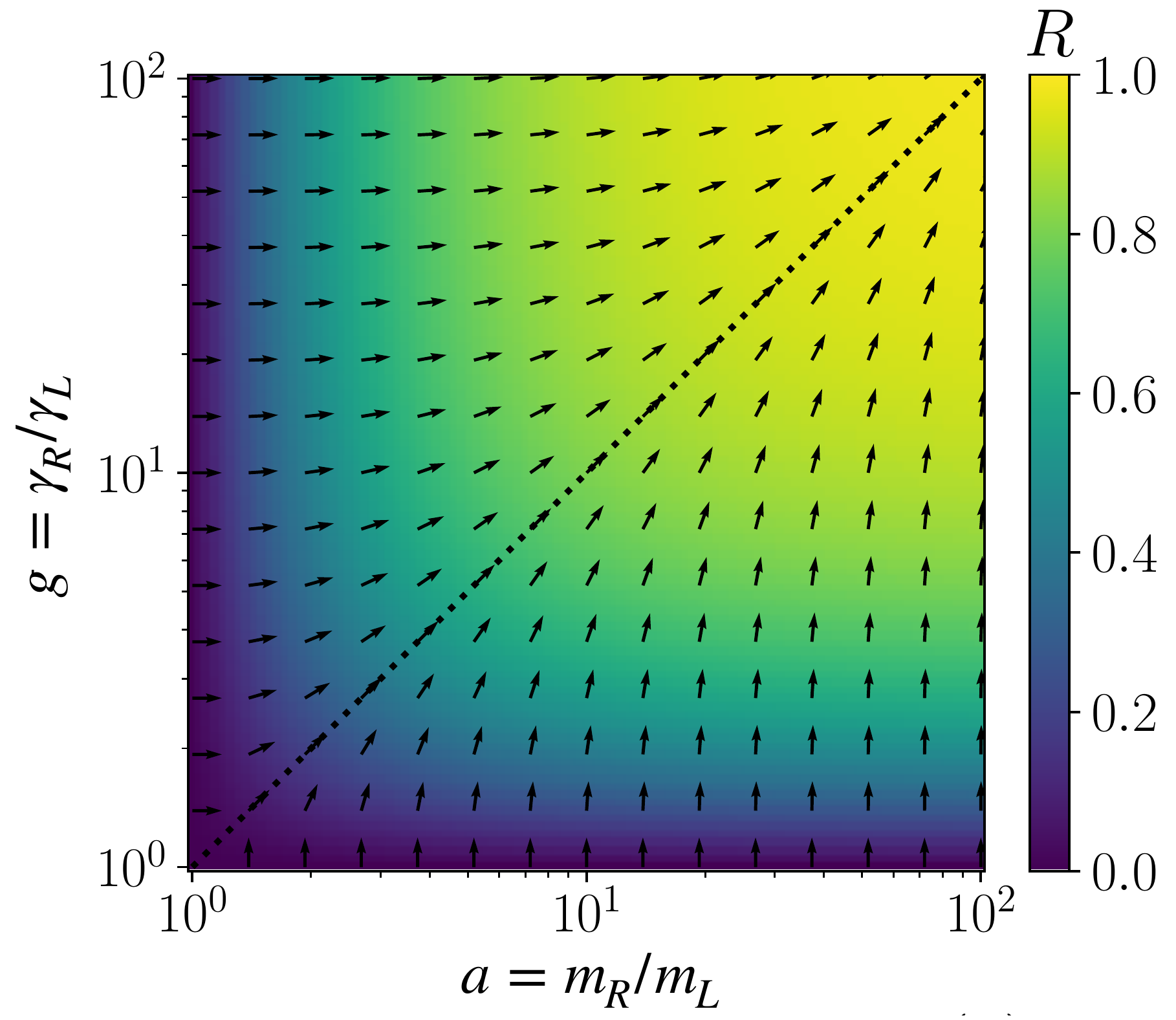}
	\caption {Rectification, $R$, given by Eq. (\ref{a10}) as a function of the ratios $a$ and $g$.
	The arrows give the gradient direction.
  }
	\label{FIGR2}
\end{figure}
%

In \cite{S2021} it was found that the region for maximal rectification for fixed masses could be
described analytically, and in the weak dissipation regime $\left(\gamma_L / m_L \ll \sqrt{k / m_L}, \ \gamma_R / m_R \ll \sqrt{k / m_R}\right)$
it is a straight line in the $k_L,k_R$ plane,
\cite{S2021}
\begin{equation}
\frac{k+k_R}{m_R}=\frac{k+k_L}{m_L}.\label{a8}
\end{equation}
On the maximum-rectification line (\ref{a8}) the rectification only depends on the mass and friction coefficient ratios $a$ and $g$,
\begin{equation}
R= \begin{cases}1-\frac{a+g}{1+a g} & \text { if } a>1, g>1 \text { or } a<1, g<1
\\
1-\frac{1+a g}{a+g} & \text { if } a>1, g<1 \text { or } a<1, g>1\end{cases},
\label{a10}
\end{equation}
where
\begin{equation}
a=m_R / m_L, \quad g=\gamma_R / \gamma_L.
\end{equation}
%
Increasing $a$ or $g$ increases the asymmetry of the system and the rectification.
From Eq. (\ref{a10}) we can represent $R$ in terms of $a$ and $g$, see Fig. \ref{FIGR2}.
The fastest way of increasing $R$ is following the diagonal dotted line $a=g$. For this reason, we shall mostly use the condition $a=g$ and sweep over the parameter $C\equiv a=g$. $R$ grows with $C$ towards one, but there are physical limitations to make these ratios arbitrarily large.
In particular changing $a$ is limited by the masses of the available ions.
%
In numerical examples and calculations hereafter we shall always fullfill Eq. (\ref{a8}) and
fix the following values:
in the forward configuration $k=1.17$ fN/m, $k_L=1$ fN/m, $\gamma_L=6.75 \times 10^{-22}$ kg/s, $m_L=24.305$ a. u. (for Mg$^+$),
whereas $\gamma_R$, $m_R$, and $k_R$
are set to satisfy chosen values of $g$ and $a$. Similarly $k_R$ is set to satisfy Eq. (\ref{a8}).
For the reverse configuration of bath temperatures we interchange the  friction coefficients,
$\tilde\gamma_L=\gamma_R$, $\tilde\gamma_R=\gamma_L$ , but the masses
and spring constants do not change with respect to the ones for the forward configuration.
The calculations of spectra using Eq. (\ref{sgen}) depend on these values and on the bath temperatures
(by the dependence on the temperature of the diffusion coefficients).
\begin{figure}
  \centering
	 \includegraphics[scale=0.44]{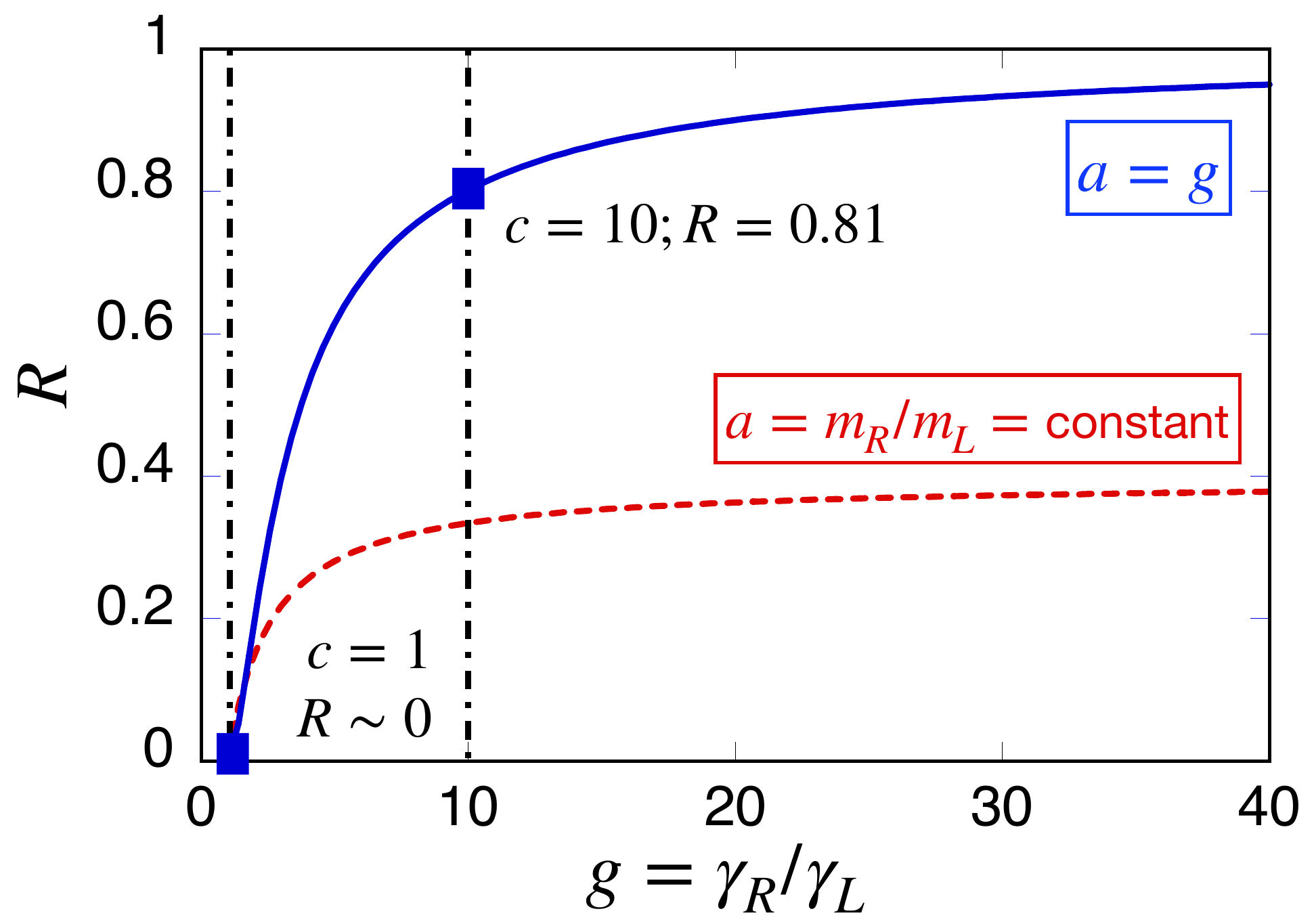}
	\caption {Rectification for different values of $g=\gamma_R / \gamma_L$. The blue solid line gives the maximal rectification,
which is  found when  for $a=g$, see Eq. (\ref{a8}). For the red dashed line the mass ratio is kept constant
(corresponding to Mg$^+$ and Ca$^+$ ions). The blue squares correspond to the values of $C=1$
(the two ions and the friction coefficients are equal, so $R\sim 0$), and $C=10, R\sim 0.8.$ The spectra for $C=10$
are depicted in Fig. \ref{FIGspec}.}
	\label{FIGR3}
\end{figure}
%

In Fig. \ref{FIGR3} the rectification is depicted versus $g$ when
$a=g$ (blue solid line), and for $a$ constant (red dashed line), which gives smaller rectification.
%
%
\subsection{Spectral densities and rectification: Example}
%
%
%
In \cite{S2021},
the spectra of the ions $S_L$ and $S_R$ for several sets of parameters exhibiting large and small rectification were studied.
Indeed the system presented large rectification if, for a bath configuration, there is a good match between the phonon spectra
of the ions and mismatch when the baths were exchanged.
\begin{figure}
	\centering
	 \includegraphics[scale=0.80]{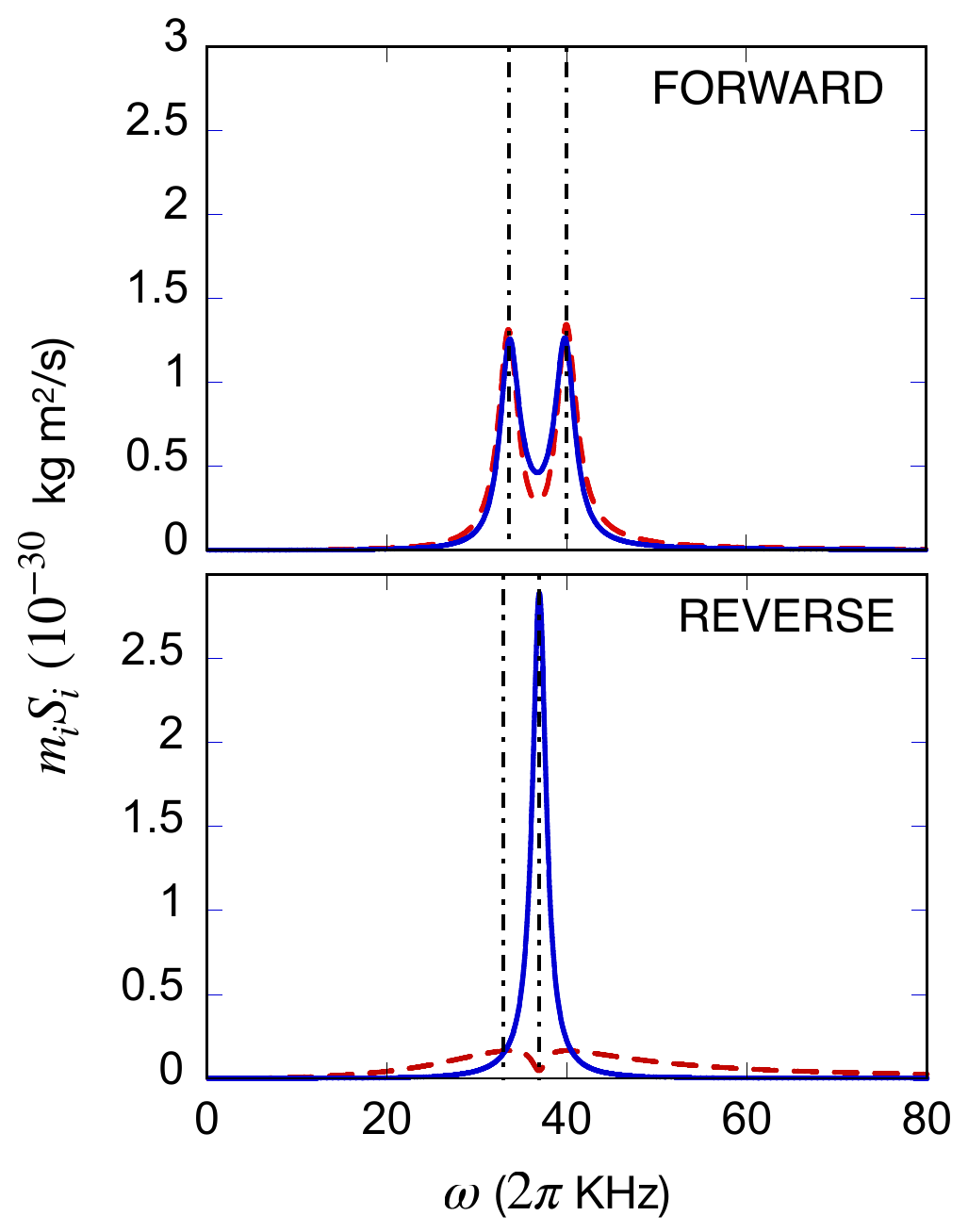}
	\caption {Spectral densities for both ions multiplied by their masses, $m_LS_L$ (solid red line) and $m_RS_R$ (dashed blue line), vs. $\omega$ for $C=10$ corresponding to forward and reverse configurations.
	$T_L=\tilde{T}_R=2 \mathrm{mK}, T_R=\tilde{T}_L=1 \mathrm{mK}$.
($\tilde{T}_i$ are the temperatures of the reverse configuration.)
	 The vertical lines are the real part of the frequencies of the dissipative normal modes of the system \cite{S2021}.
	 The areas are proportional to the particle kinetic energies (or temperatures). The rectification coefficient is
	 $R \approx 0.8$: in the forward configuration, the spectra match well while in the reverse configuration there is a clear mismatching.}
	\label{FIGspec}
\end{figure}


Figure \ref{FIGspec} shows the spectra $m_LS_L$ and $m_RS_R$ for $C=10$, i.e., for high rectification,
$R\sim 0.8$, see Fig. \ref{FIGR3}.  For forward bias there is almost perfect matching between the spectral densities
but a mismatch for the reverse configuration.
This is a clear example of the qualitative relation between flux and spectral matching.
In the following section we shall give this relation a more quantitative form.

%
%
\section{Relations between spectral matching and heat flux\label{sec:FluxMatching}}
%
%
%
%
%
%
The matching $M$ or overlap between the spectral densities has to be defined.
 A relevant definition would be one related to the flux, by a direct dependence, or by an inequality.
We may expect as an ansatz a form depending on the product of the spectra,
%
\begin{equation}
    M=\int F[S_L(\omega)S_R(\omega);\omega]\: d\omega.
    \label{match}
\end{equation}
The following discussion will find a natural, simple choice for the function $F$.

We need to average over realizations of the noise. First, we define
$X_{Rj}(\omega)$ and $X_{Lj}(\omega)$ as the Fourier transforms of the  displacements, $q_{Lj}(t)$ and $q_{Rj}(t)$
respectively, in the $j$-th realization. We separate real and imaginary parts,
\begin{eqnarray}
    X_{Rj}(\omega)&=&a_j+b_ji,
    \nonumber\\
    X_{Lj}(\omega)&=&c_j+d_ji.
\end{eqnarray}
Notice that $X_{Lj}(-\omega)=c_j-d_ji$ because the displacements are real. Therefore,
\begin{eqnarray}
    \langle X_R(\omega)  X_L(-\omega)\rangle
    \!=\!\sum_{j}^N\! \frac{a_jc_j+b_jd_j+i(c_jb_j-a_jd_j)}{N},
    \label{18}
\end{eqnarray}
where $N$ is the number of realizations which is supposed to be very large. We are only interested in the imaginary part of (\ref{18})
according to the flux expression (\ref{elflujo1}). The square of the imaginary part is
%
\begin{eqnarray}
&&\hspace*{-.7cm}(c_jb_j-a_jd_j)^2=a_j^2d_j^2+b_j^2c_j^2-2a_jc_jb_jd_j
\nonumber\\
&&\hspace*{-.7cm}\leq a_j^2c_j^2+b_j^2d_j^2+b_j^2c_j^2+a_j^2d_j^2=(a_j^2+b_j^2)(c_j^2+d_j^2).
\end{eqnarray}
%
From this inequality we conclude that
\begin{align}
\begin{split}
    |(c_jb_j-a_jd_j)|\leq  \sqrt{a_j^2+b_j^2}\sqrt{c_j^2+d_j^2}.
    \label{20}
\end{split}
\end{align}
The absolute value of the imaginary part of the correlation function in Eq. (\ref{18}) is
\begin{align}
\begin{split}
    |\operatorname{Im}\left[ \langle X_R(\omega) X_L(-\omega)\rangle \right]|=\left| \sum_{j}^N \frac{c_jb_j-a_jd_j}{N}\right|
\end{split}
\end{align}
and, from Eq. (\ref{20}), it obeys
\begin{equation}
   \left| \sum_{j}^N \frac{c_jb_j-a_jd_j}{N}\right|\leq\sum_{j}^N \frac{\sqrt{a_j^2+b_j^2}\sqrt{c_j^2+d_j^2}}{N}.
   \label{k1}
\end{equation}
Now, we apply the Cauchy-Bunyakovsky-Schwarz inequality,
\begin{equation}
\left(\sum_{j=1}^{n} \alpha_{j} \beta_{j}\right)^{2} \leq\left(\sum_{j=1}^{n} \alpha_{j}^{2}\right)\left(\sum_{j=1}^{n} \beta_{j}^{2}\right),
\end{equation}
%
%
%
to the right-hand side of (\ref{k1}), with $ \sqrt{a_j^2+b_j^2}=\alpha_j$ and $ \sqrt{c_j^2+d_j^2}=\beta_j$, to find
\begin{eqnarray}
  &&\sum_{j}^N \frac{\sqrt{a_j^2+b_j^2}\sqrt{c_j^2+d_j^2}}{N}  \leq
  \nonumber\\
  &&\frac{1}{N}\sqrt{\left( \sum_{j}^N a_j^2+b_j^2\right)\left( \sum_{j}^N c_j^2+d_j^2\right)}.
  \label{k2}
\end{eqnarray}
As
\begin{align}
\begin{split}
  \langle X_R(\omega)X_R(-\omega)\rangle=\frac{1}{N}\sum_{j}^N (a_j^2+b_j^2),\\
  \langle X_L(\omega)X_L(-\omega)\rangle=\frac{1}{N}\sum_{j}^N (c_j^2+d_j^2),
\end{split}
\end{align}
 we find from Eq. (\ref{k2}) the following inequality for the integrand in Eq. (\ref{elflujo1}),
%
\begin{eqnarray}
&&\left| \operatorname{Im} \left[ \langle X_R(\omega) X_L(-\omega)\rangle \right]\right|
\nonumber\\
&&\leq\sqrt{ \langle X_L(\omega)X_L(-\omega)\rangle\langle X_R(\omega)X_R(-\omega)\rangle}.
 \label{k4}
\end{eqnarray}
%
Since $\langle X_i(\omega)X_i(-\omega)\rangle$ is related to $S_i(\omega)=\langle W_i(\omega)W_i(-\omega)\rangle$
by Eq. (\ref{Ss}),
%
%
expression (\ref{k4}) can be written as
\begin{equation}
 \omega^2\left| \operatorname{Im} \left[ \langle X_R(\omega) X_L(-\omega)\rangle \right] \right| \leq\sqrt{ S_L(\omega)S_R(\omega)},
\end{equation}
or
%
%
using Eq. (\ref{elflujo1}),
%
%
and taking into account that $|\int f(x)dx| \leq \int |f(x)| dx$,
%
\begin{align}
\label{matching}
\begin{split}
 |J| \leq \frac{k}{2\pi} \int \frac{1}{|\omega|}\sqrt{ S_L(\omega)S_R(\omega)}\: d\omega,
\end{split}
\end{align}
which sets an upper limit for the heat flux.
This relation prompts us to define the function $F$ in Eq. (\ref{match}) and the matching as
%
\begin{align}
\label{matching1}
\begin{split}
 M= \frac{k}{2\pi} \int \frac{1}{|\omega|}\sqrt{ S_L(\omega)S_R(\omega)}\: d\omega.
\end{split}
\end{align}
%
This measure of the matching (\ref{matching1}) allows for direct comparison between $J$ and $M$ since they have the same dimensions while with other proposed definitions we can only compare their ratios \cite{Li}. When so defined, the spectral density matching sets an upper bound for the flux,
$|J|\leq M$. 

In \cite{ITRLi} Li {\it{et al.}}, to quantify the overlap between the power spectra between
left and right segments, introduced
\begin{equation}
{\cal S}=\frac{\int_0^{\infty} S_L(\omega) S_R(\omega) d w}{\int_0^{\infty} S_L(\omega) d w \int_0^{\infty} S_R(\omega) d w}.
\end{equation}
and demonstrated the correlation between the heat fluxes and the overlaps of the spectra.
They found numerically the relation $|J/\tilde{J}|^\delta \sim ({\cal S}/{\tilde{\cal S}})$, with $\delta=1.62\pm 0.10$, in their model, two weakly linearly coupled, dissimilar anharmonic segments, exemplified by a Frenkel-Kontorova chain segment and a neighboring Fermi-Pasta-Ulam chain segment.

Next, we will evaluate the flux and the matching for different parameter configurations for the two-ion model to test the inequality $|J|\leq M$ and also to look
for a similar relation to the one found by Li {\it{et al.}}  but for the matching expression introduced here.
%
%
%
\section{Flux and matching for the two-ion model \label{sec:NumRes}}
%
%
%
%
We compute the heat flux and the matching (\ref{matching1}) for the two-ion model solving Eqs. (\ref{Js}) and (\ref{a7}) for different parameter configurations. We will only consider the maximal rectification region given by the condition (\ref{a8}).

\begin{figure}
	\centering
 \includegraphics[scale=0.43]{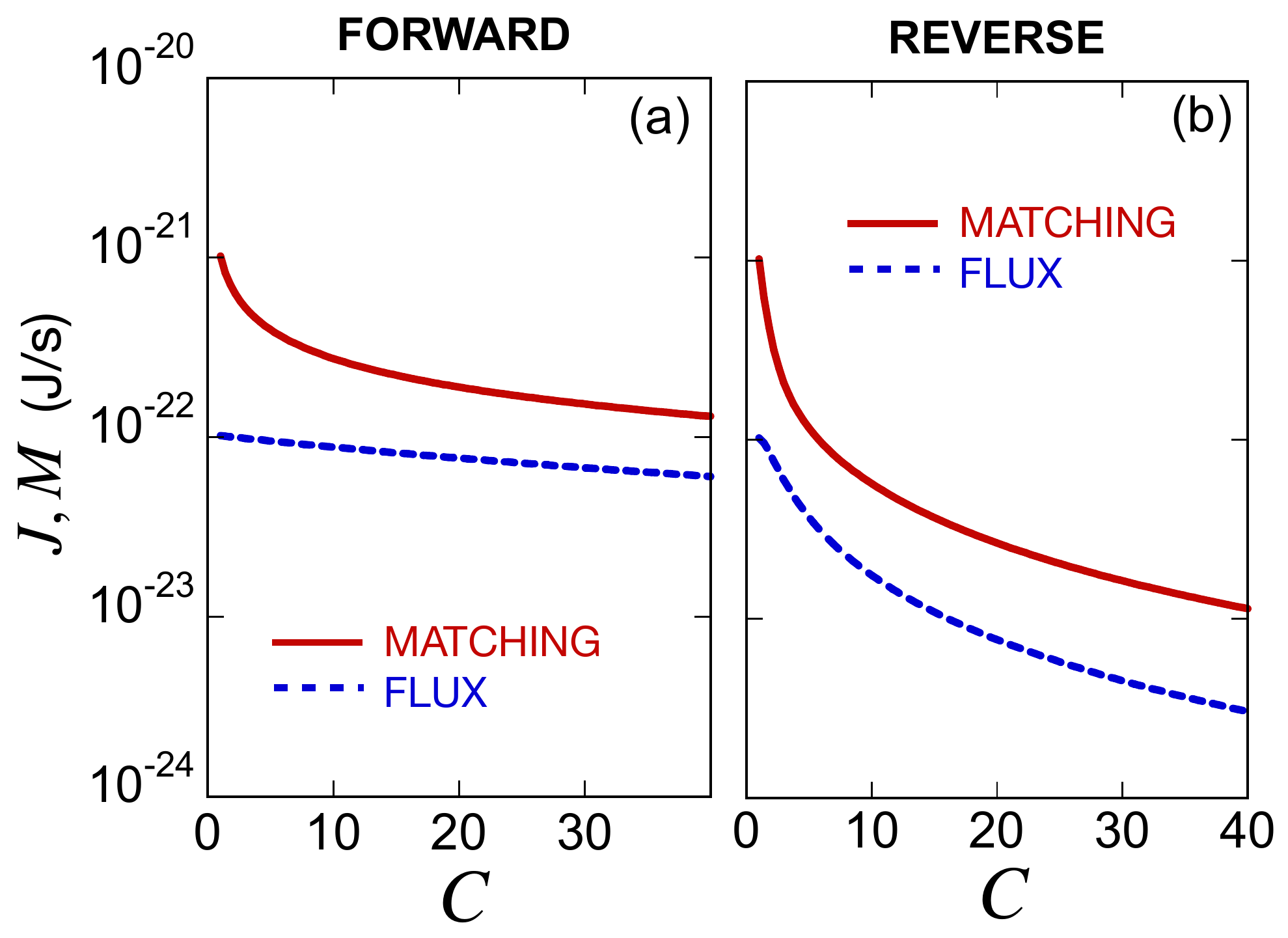}
	\caption {Flux and matching  versus  $C=a=g$.
$T_L=1$ mK, $T_R=0.1$ mK. Other parameters, as explained in the Sec. \ref{recti}.}
	\label{FIGJM1b}
\end{figure}

In Fig. \ref{FIGJM1b} the flux and the matching are displayed as a function of $C=g=a$.
As predicted by Eq. (\ref{matching}) the matching is above the flux. Both quantities behave similarly and the difference tends to a constant as $C$ increases. Also, $R$ tends to one, as seen in Fig. \ref{FIGR3}.
The forward and reverse configurations show very different curves (a sign of rectification)
as we are following the line of fastest growth of $R$.

\begin{figure}
	\centering
	 \includegraphics[scale=0.43]{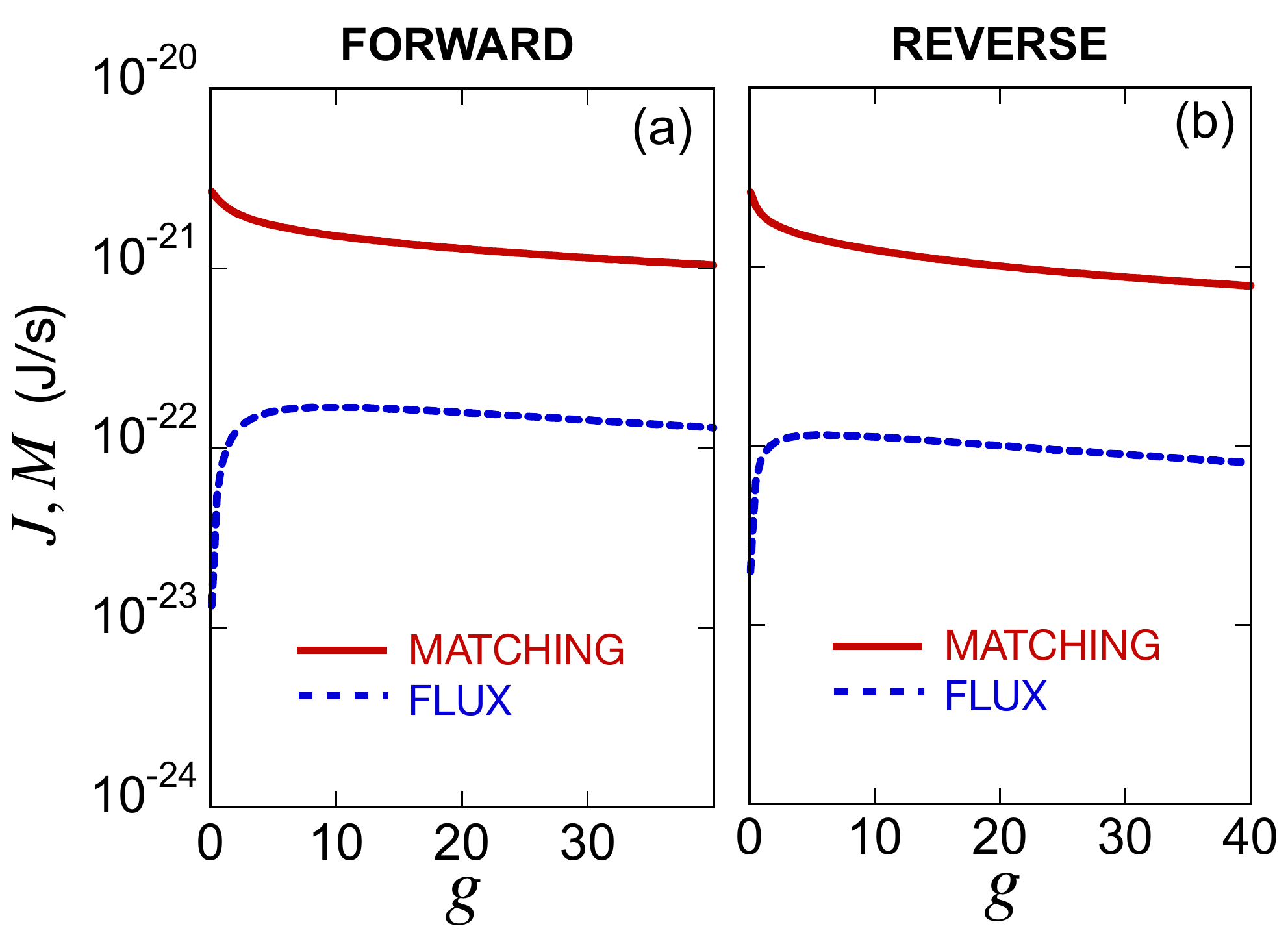}
	\caption {Flux and matching versus $g$ with a constant mass ratio $a$.
$T_L=1$ mK, $T_R=0.1$ mK. Other parameters, as explained in the Sec. \ref{recti}.}
	\label{FIGJM2b}
\end{figure}

Since experimentally it is not feasible to have a continuum for the masses ratio $a$, in Fig. \ref{FIGJM2b} we have also plotted $J$ and $M$ fixing the masses for Ca$^+$ and Mg$^+$ ions, and sweeping over $g$. Both quantities behave similarly, except in the region for very low $g$, but this region  is not really interesting since it corresponds to a very low $R$. Forward and reverse curves are now closer to each other, corresponding to a smaller $R$, see again Fig. \ref{FIGR3}.


\begin{figure}
	\centering
	 \includegraphics[scale=0.57]{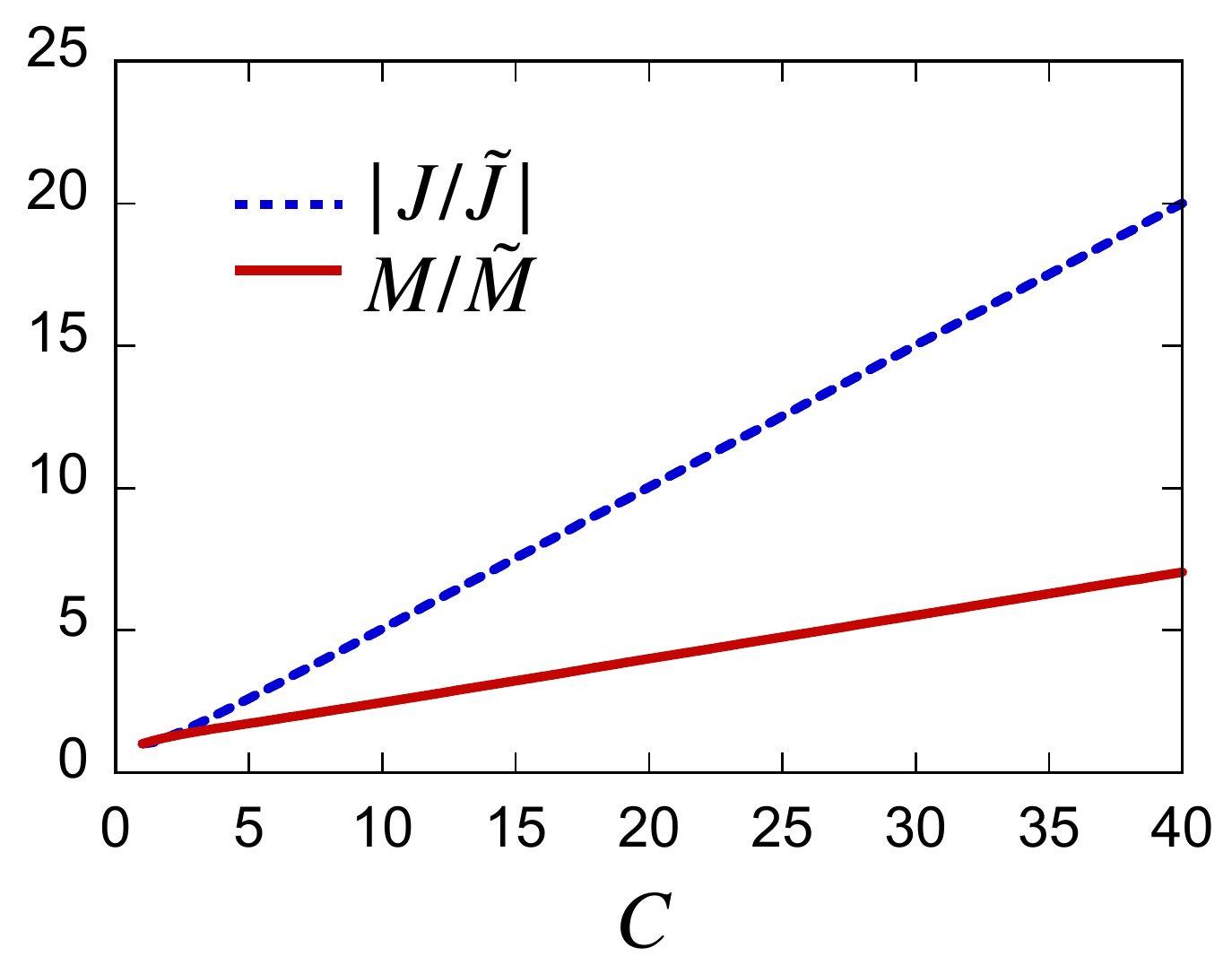}
	\caption{Forward to reverse ratios $J/\tilde{J}$ and $M/\tilde{M}$ versus $C$.  $T_L=1$ mK, $T_R=0.1$ mK}
	\label{ratios}
\end{figure}

The flux ratio $J/\tilde{J}$ is exactly linear with $C$ according to Eq. (\ref{rdefi}), see Fig. \ref{ratios},
whereas the ratio $M/\tilde M$ is also linear in $C$, except for very low $C$, with a proportionality factor that depends on the
ratio of the bath temperatures.

\begin{figure}
	\centering
	 \includegraphics[scale=0.37]{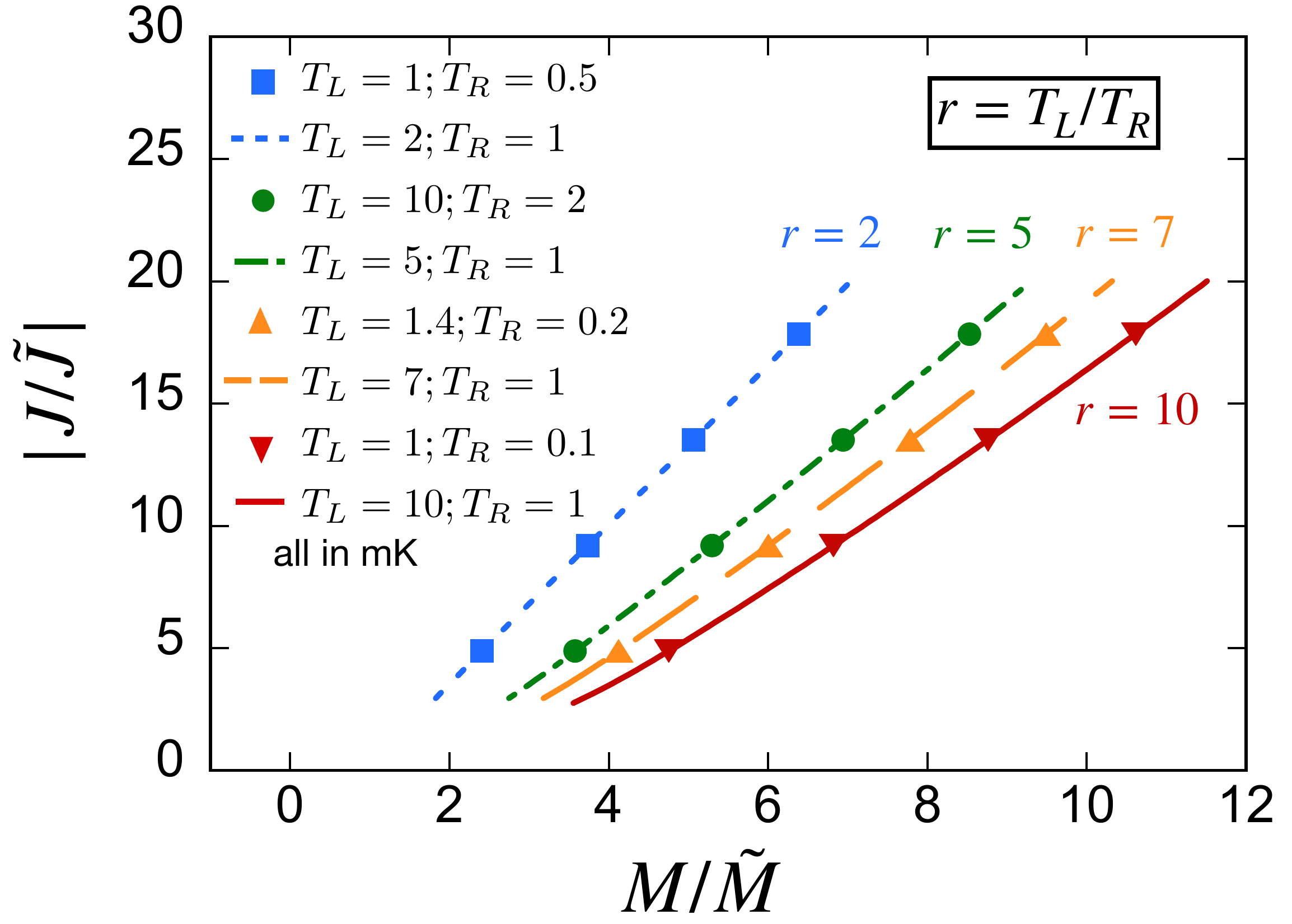}
	\caption {Flux ratio $J/\tilde{J}$ vs. the matching ratio $M/\tilde{M}$ for different temperature intervals.  The line points depend parametrically on $C$.  The slope depends on the ratio of temperatures $T_L/T_R$. Dots and lines of the same color indicate same temperatures' ratio but different values of the temperatures (in mK) as indicated in the legend. For instance, blue dots correspond to $T_L=1$ mK and $T_R=0.5$ mK and the blue line to $T_L=2$ mK and $T_R=1$ mK, both cases verify $T_L/T_R=2$. }
	\label{FIGMsJsdiffT}
\end{figure}

Thus, sweeping over $C=a=g$ a very simple linear relation is found numerically between  the ratios $J/\tilde{J}$ and $M/\tilde{M}$ for our model,
%
\begin{align}
\label{slope}
\begin{split}
  |J/\tilde{J}| \sim M/\tilde{M},
\end{split}
\end{align}
with a proportionality factor that depends on the ratio between temperatures $T_L/T_R$, as shown in Fig. \ref{FIGMsJsdiffT}.

%
%
%
%
\section{Discussion\label{sec:conclusions}}
Using a simple, but experimentally feasible  model of two ions interacting with laser-induced heat baths, we have defined the power spectrum overlap or
``spectral matching'' of the ions so that it provides an upper bound to the flux. In fact forward to reverse flux ratios are proportional to matching ratios
for the parameter conditions where rectification is optimal. These findings put on a sounder basis the relation between heat rectification and the spectral match or mismatch for forward and reverse bath temperatures.

\begin{figure}[!t]
\centering
\vspace*{-6.5cm}
\hspace*{-1.5cm}
\includegraphics[scale=0.55]{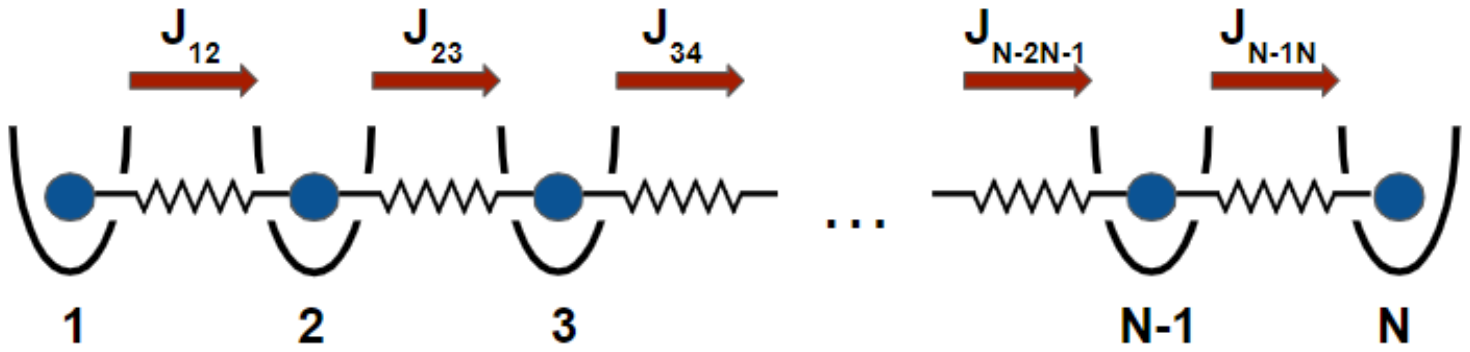}
\vspace*{-7cm}
\caption {$N$-particle linear chain with interaction between nearest neighbors.}
\label{FGRNiones}
\end{figure}

The results can be generalized to any $N$-particle chain with linear interactions between nearest neighbors
and two thermal baths at the boundaries. The trap potentials could be anharmonic.
For the $N$-particle chain in the steady state $\langle \dot{H}_i\rangle=0$, see Fig. \ref{FGRNiones},
where $H_i$ is the local energy for the $i$th particle. Thus the energy flux from particle $i-1$ to particle $i$ equals the flux from
particle $i$ to particle $i+1$, namely  $J_{i-1,i}=J_{i,i+1}$.

The flux that crosses the  chain $J$ is
\begin{equation}
    J=J_{i,i+1}
\end{equation}
where $i$ can be $1,2,3,...,N-1$.
The equations from (\ref{eq13}) to (\ref{elflujo1}) and the arguments in Sec. \ref{sec:FluxMatching} for ions $L$ and $R$ are valid as well for particles $i$ and $i+1$, so
\begin{equation}
    |J_{i,i+1}|\leq M_{i,i+1},
\end{equation}
where $M_{i,i+1}$ is the matching (\ref{matching1}) between the spectral densities of ions $i$ and $i+1$.
 Therefore
\begin{equation}
    |J|\leq M_{i,i+1},
    \label{fr}
\end{equation}
where $i$ can be $1,2,3,...,N-1$.
Eq. (\ref{fr}) is the generalization of our results for an $N$-particle linear chain and it states that the flux through the chain is
bounded by the spectral matching of nearest-neighbor particles.

\begin{acknowledgments}
	We thank Miguel \'Angel  Sim\'on for useful discussions.
	We thank the Grant PID2021-126273NB-I00 funded by MCIN/AEI/ 10.13039/501100011033 and by “ERDF A way of making Europe”.
	We acknowledge financial support from the Basque Government Grant No. IT1470-22. MP acknowledges support from the Spanish Agencia Estatal de Investigación, Grant No. PID2019- 107609GB-I00.
\end{acknowledgments}


\bibliography{apssamp}
\end{document}